\documentclass[final,twoside,11pt]{entics} 
\usepackage{enticsmacro}

\input{packages}

\newcommand{\undf}[1]{#1}

\newcommand{\To}{\mathbin{\Rightarrow}}
\newcommand{\commentout}[1]{}

\newcommand{\suplus}{\mathrel{\uplus}}

\newcommand{\partsof}[1]{\pow{#1}}
\newcommand{\pow}[1]{2^{#1}}
\newcommand{\Nat}{\mathbb N}
\newcommand{\sleq}{\mathbin{\sqsubseteq}}
\newcommand{\ttt}{\mathrm{t\!t}}
\newcommand{\fff}{\mathrm{ff}}
\newcommand{\dvg}{{\bot\!\!\!\!\bot}} 
\newcommand{\fsp}[2]{[#1 \to #2]}
\newcommand{\deq}{\mathrel{\coloneqq}}
\newcommand{\powo}[1]{\undf{\Omega^{#1}}}

\newcommand{\fa}[1]{\forall #1~.~}
\newcommand{\ex}[1]{\exists #1~.~}

\newcommand{\lam}[1]{\lambda #1~.~}

\newcommand{\Pos}{\mathbf{Pos}}

\newcommand{\Rel}{\mathbf{Rel}}
\newcommand{\Set}{\mathbf{Set}}

\newcommand{\CLat}{\mathbf{CLat}}
\newcommand{\Aff}{\mathbf{Aff}}
\newcommand{\real}{\mathbbm{R}}

\newcommand{\lwhile}{L_{\mathrm{w}}}
\newcommand{\lwhilev}{L_{\mathrm{wv}}}

\newcommand{\cpow}{Q} 

\newcommand{\mpow}{P}
\newcommand{\mmay}{M}
\newcommand{\msdist}{D_s}

\newcommand{\sem}[1]{\llbracket #1\rrbracket}
\newcommand{\asem}[1]{\llparenthesis #1\rrparenthesis}
\newcommand{\csem}[1]{\llbracket #1\rrbracket_c}

\newcommand{\semt}[2][T]{\llbracket #2\rrbracket_{#1}}
\newcommand{\psem}[1]{\llparenthesis #1\rrparenthesis}

\newcommand{\dref}[1]{Definition~\ref{d:#1}\xspace}
\newcommand{\eref}[1]{Example~\ref{e:#1}\xspace}

\newcommand{\sref}[1]{Section~\ref{s:#1}\xspace}
\newcommand{\tref}[1]{Theorem~\ref{t:#1}\xspace}

\newcommand{\nt}[1]{\{\text{#1}\}}

\newcommand{\Obj}{\mathrm{Obj}}
\newcommand{\id}{\mathrm{id}}
\newcommand{\tmop}[1]{\ensuremath{\operatorname{#1}}}

\newcommand{\adjdashv}[2]{
  \ar@<#2>@{}[#1]|-*=0[@]\txt{\tiny{$\bot$}} 
}
\newcommand{\adjunction}[3]{
  \ar@<.3pc>[#1]^-{#2}
  \adjdashv{#1}{0pc}
  \ar@<-.3pc>@{<-}[#1]_-{#3}
}

\newcommand{\funcdefi}[4]{$#1\deq #2$ and $#3\deq #4$}

\newcommand{\galois}[4]{#1\dashv #2:#3\to #4}

\newcommand{\scr}[2]{#2_{#1}}
\newcommand{\timespos}{\scr\Pos\times}
\newcommand{\Topos}{\scr\Pos\To}

\newcommand{\propfun}{\textcolor{black}{property functor}}

\newcommand{\PropFun}{\textcolor{black}{Property Functor}}
\newcommand{\propfuns}{\textcolor{black}{\propfun s}}

\newcommand{\intp}{\textcolor{black}{interpretation}}

\newcommand{\intps}{\textcolor{black}{\intp s}}

\newcommand{\gcs}{functorial collecting semantics}
\newcommand{\GCS}{Functorial Collecting Semantics}

\newcommand{\concd}{\textcolor{black}{concretization of domains}}
\newcommand{\conci}{\textcolor{black}{concretization of interpretation}}
\newcommand{\absd}{\textcolor{black}{abstraction of domains}}
\newcommand{\absi}{\textcolor{black}{abstraction of interpretation}}

\newcommand{\kwd}[1]{\mathbf{#1}}
\newcommand{\kwelse}{\kwd{else}}
\newcommand{\kwif}{\kwd{if}}
\newcommand{\kwskip}{\kwd{skip}}
\newcommand{\kwwhile}{\kwd{while}}
\newcommand{\kwaddvar}{\kwd{addvar}}
\newcommand{\kwdelvar}{\kwd{delvar}}

\newcommand{\wasgn}[2]{#1 := #2}
\newcommand{\wcond}[3]{\kwif\, #1\,\{#2\}\kwelse\{#3\}}
\newcommand{\wwhile}[2]{\kwwhile\, #1\,\{#2\}}

\newcommand{\grd}[2]{\undf{[#1=#2]}}

\newcommand{\svars}{X}
\newcommand{\svals}{\bbV}
\newcommand{\smems}{\bbM}
\newcommand{\smemsbot}{\smems_{\dvg}}

\newcommand{\amem}{m^{\sharp}}

\newcommand{\Abs}[1]{\text{Abs}(#1)}

\newcommand{\qpow}[1]{Q(#1)}

\newcommand{\mathboldcommand}[1]{\mathbb{#1}}

\newcommand{\bbM}{\mathboldcommand{M}}

\newcommand{\bbV}{\mathboldcommand{V}}

\usepackage{euscript}
\newcommand{\mathcalcommand}[1]{\mathcal{#1}}

\newcommand{\mcC}{\mathcalcommand{C}}

\newcommand{\mcE}{\mathcalcommand{E}}

\newcommand{\mcL}{\mathcalcommand{L}}

\DeclareMathAlphabet{\mathpzc}{T1}{pzc}{m}{it}

\newcommand{\confi}[1]{}

\def\conf{MFPS 2023} 	
\def\myconf{mfps39}
\volume{3}			
\def\volu{3}			
\def\papno{11}				
\def\lastname{Katsumata, Rival, Dubut} 
\def\runtitle{Categorical Abstraction} 
\def\mydoi{12288}
\def\copynames{S. Katsumata, X. Rival, J. Dubut}    

\begin{document}
\newcommand{\typesetemail}[1]{\href{mailto:#1} {\texttt{\normalshape #1}}}
\newcommand{\fillhere}{\undf{*** Fill Here **}}
\begin{frontmatter}
  \title{A Categorical Framework for Program Semantics
    and Semantic Abstraction}

  \author{Shin-ya Katsumata\thanksref{b}\thanksref{email-shinya}}
  \address[b]{National Institute of Informatics}
  \thanks[email-shinya]{Email: \typesetemail{s-katsumata@nii.ac.jp}}

  \author{Xavier Rival\thanksref{c}\thanksref{email-xavier}}
  \address[c]{INRIA Paris, Département d'Informatique de l'Ecole Normale Supérieure de Paris, Université PSL, CNRS}
  \thanks[email-xavier]{Email: \typesetemail{rival@di.ens.fr}}

  \author{J\'er\'emy Dubut\thanksref{a}\thanksref{email-jeremy}}
  \address[a]{National Institute of Advanced Industrial Science and Technology}
  \thanks[email-jeremy]{Email: \typesetemail{jeremy.dubut@aist.go.jp}}
  
  \begin{abstract}
    Categorical semantics of type theories are often characterized as
    structure-preserving functors.
    This is because in category theory both the syntax and the domain of
    interpretation are uniformly treated as structured categories, so that
    we can express interpretations as structure-preserving functors between
    them.
    This mathematical characterization of semantics makes it convenient to
    manipulate and to reason about relationships between interpretations.
    Motivated by this success of functorial semantics, we address the
    question of finding a functorial analogue in {\em abstract
      interpretation}, a general framework for comparing semantics, so
    that we can bring similar benefits of functorial semantics to
    semantic abstractions used in abstract interpretation.
    Major differences concern the notion of interpretation that is being
    considered.
    Indeed, conventional semantics are value-based whereas abstract
    interpretation typically deals with more complex properties.
    In this paper, we propose a functorial approach to abstract
    interpretation and study associated fundamental concepts therein.
    In our approach, interpretations are expressed as oplax functors in
    the category of posets, and abstraction relations between interpretations
    are expressed as lax natural transformations representing
    concretizations. We present examples of these formal concepts from
    monadic semantics of programming languages and discuss soundness.
  \end{abstract}
\end{frontmatter}

\section{Introduction}
In a categorical setting, programs semantics can often be
characterized as functors.
In this setup, programs are viewed as morphisms and morphism composition
describes how programs can be composed.
As an example, in the case of typed functional programs, one may let
objects be types and morphisms be functions.
More precisely, a morphism from object $a$ to object $b$ denotes a
function of type $a \rightarrow b$.
Then, the semantics maps programs to morphisms between objects that
interpret input and output elements into a well-chosen semantic domain.
This construction is very general and accepts a wide range of semantic
domains.

This functorial presentation of semantics is prominent in the
categorical semantics of type theories and algebraic theories. There,
both type theories and semantic categories are treated as categories
possessing a common structure, and the semantics itself is presented as
a structure-preserving functor.
A typical example is the categorical semantics of the simply typed
lambda-calculus, commonly studied under the well-known Curry-Howard-Lambek
correspondence \cite{ls:cup88}; the calculus
modulo $\beta\eta$-equality is presented as a Cartesian closed
category, and its semantics in a Cartesian closed category is
presented as a functor preserving finite products and
exponentials. Another example is the categorical presentation of
algebraic theories as a particular kind of categories with finite
products (called {\em Lawvere theories} \cite{l:th63,arv:cup10}), and
their models as finite-product preserving functors.
These categorical and syntax-free presentations of the calculus and its
semantics brought significant convenience and advances to the study of
type theories and their semantics.
Additionally, monads turned out to be the tool of choice in order to
construct semantics for effectful programs~\cite{m:ic:91}. 

{\em Abstract interpretation}~\cite{cc:popl:77} provides a framework to
compare program semantics of varying levels of expressiveness, and to
derive sound approximations of program semantics, based on a given
abstraction relation.  It has been used to describe relationships
across program semantics~\cite{pc:mfps:97}, program
analysis~\cite{cc:popl:77,astree:pldi:03,fb:cacm:19}, program
transformations~\cite{cc:popl:02}, and more.  However, it is usually
formalized in order 
theory since this presentation suffices in many applications.
Therefore, it is not immediately compatible with the aforementioned
categorical presentation.

Although the notion of Galois connection, which is abundantly used in
abstract interpretation works~\cite{cc:popl:77,cc:jlc:92}, is adjunctions
between posets, few works have studied a more complete
description of abstract interpretation frameworks in a categorical setup.
Among the works that relied on categorical tools in order to describe some
specific semantic abstraction concepts for specific purposes, we can
cite, Steffen et al.~\cite{sjm:tia:92} who integrate both concrete
and abstract semantics in a categorical settings in order to examine
questions related to soundness and completeness, with respect to a
given set of behaviors.
Venet~\cite{av:sas:96} uses mathematical tools that stem from category
theory in order to construct specific families of abstract domains.
More precisely, he applies the Grothendieck construction to generalize
constructions such as reduced product and cardinal power~\cite{cc:popl:79}.
More recently, Sergey et al.~\cite{might:pldi:13} took advantage of the
monadic structure of a semantics of lambda-calculus to derive a static
control flow analysis for a small functional language as well as an
implementation in Haskell.

In this work, we seek for more comprehensive foundations for
classical abstract interpretation techniques into the categorical
semantics settings.
We start with an interpretation of programs as morphisms in a syntactic
category and semantics as functors from programs to the category of
posets.
We formalize and generalize the notion of collecting semantics typically
used in program analysis as a decomposition of such a functor.
In this setup, we integrate the notion of abstraction using some form
of natural transformations between these functors.
More precisely, the approximation inherent in sound, incomplete
abstractions are accounted for using lax natural transformations.
We show that this construction also enables the abstract interpretation
of a basic language.

To achieve these goals, we build upon a categorical interpretation of
programs and their semantics.  In our categorical formalism, the
design of an abstract semantics with respect to a denotational
semantics $\sem-:L\to\mcC$ proceeds as follows.
First, we turn the denotational semantics into a \emph{functorial
collecting semantics} by composing $\sem{-}$ with a functor $C:\mcC\to\Pos$
(where $\Pos$ denotes the
category of posets and monotone functions between them), which we call
\emph{property functor}.  This functor plays the role of attaching a
notion of property and a direct image operation to the category
$\mcC$.  This step is crucial for the design of the analysis, as it
fixes the concrete semantics the analysis is built upon.  Then, an
analysis using abstract domains over the collecting
semantics $C\circ\sem-$ is expressed as an {\em oplax} functor
$A:L\to\Pos$ equipped with a {\em lax} natural transformation
$\gamma:A\to C\circ\sem-$ representing a {\em \conci{}}:
\begin{equation}
  \label{eq:abs}
  \xymatrix@R=.3cm@C=2cm{
    L \ar[rd]_-{\sem-} \ar@/^.5cm/[rr]^-{A} \ar@{}[rr]|-{\Downarrow\gamma}
    & & \Pos \\
    & \mcC \ar[ru]_-{C}
  }
\end{equation}
Here, the functor $A$ being oplax means that it only satisfies
weakened functor axioms. The lax natural transformations are also
weakening of natural transformations, replacing the naturality axiom
to an inequality. The use of oplax functors for modeling analysis was
initiated by Steffen, Jay and Mendler \cite{sjm:tia:92}.  We adopt the
same approach, and further bring some basic concepts that are not
covered in \cite{sjm:tia:92} into the oplax functor formalism.

The common approach relies on fixing such an abstraction relation
(here described by $\gamma$) and seeking for a sound, possibly
approximate $A$ that can be implemented efficiently.  A natural and
important question is how such an $(A,\gamma)$ pair can be
constructed.  This can be done by extending the collecting semantics
with a family of {\em Galois connections}, which are abundantly used
in abstract interpretation, or with a family of concretization
functions when best abstraction cannot be ensured.

This story naturally extends to the monadic semantics of various
effectful programming languages.  Indeed, as discussed earlier, the
semantics of such programs is often derived using Kleisli categories
of monads.  Assuming a base category $\mcC$ for values and a monad $T$
for effects, effectful programs are interpreted in the Kleisli category
$\mcC_T$, and the semantics takes the form of a functor $F: L\to \mcC_T$.
We then derive a {\em collecting semantics} by composing it with a
functor $\mcC_T\to\Pos$, and its abstraction is given, following the
lax natural transformation discussed previously.  Therefore, another
benefit of our approach is to simplify the design of static analyses
for effectful programs, thanks to a better integration of program
semantics and abstraction.

To summarize, upon the work by Steffen, Jay and Mendler
\cite{sjm:tia:92}, we formalize abstract interpretation in a
functorial semantics framework.  The new ingredients from
\cite{sjm:tia:92} are the following:
\begin{enumerate}
\item We show that interpretations (oplax functors) are closed under
  the {\em induction operation} by Galois connections (\tref{galcon}).  This induction
  also comes with {\em concretizations of interpretations},
  formulated as lax natural transformations.
  
\item We give a categorical formulation of {\em collecting semantics} (\sref{3:colsem}),
  which is the starting point of the development of abstract
  interpretations. In our formulation, a collecting semantics is an
  extension of a standard denotational semantics with a {\em property
    functor}, which attaches forward predicate transformers to
  the model category of the denotational semantics.

\item We present two examples of developments of abstract
  interpretations, one for a while language over generic computational
  effects (\sref{3:colsem} and \sref{4:abs}), and the other for the
  simply typed lambda calculus in \sref{5:systemt}.
  An additional result that follows from this approach 
  is a strongest
  postcondition predicate transformer semantics for the while language
  over general monads and truth value complete lattices
  (\tref{inductwhile}). This semantics is a generalization of the
  strongest postcondition semantics introduced in \cite{10.1145/3527331}.
\end{enumerate}


\section{Interpretation as an Oplax Functor}
\label{s:2:oplax}
In this section, we set up basic definitions that serve as foundations
for our integrated framework in the next sections.
At this point, our main goal is to formalize the construction of semantics
from programs.

\subsection{Programs}
\label{s:2:1:progs}
Before considering semantic interpretations, we need to fix syntactic
definitions.
In this paper, we make the choice to also integrate programs as categorical
entities, so that we can define interpretations as functors.
This presentation is natural since programs can be viewed as transformations
from inputs to inputs, that can also be composed like morphisms in a
category.
To do that, we follow a classical idea that lets a category $\mcL$ stand
for the syntactic definition of programs and their inputs/outputs;
more precisely,
\begin{compactitem}
\item objects describe elements consumed/produced by programs;
\item programs stand for morphisms.
\end{compactitem}
As an example, in a functional setup, we may let objects be types (thus
describing values) and morphisms be functions mapping values to values.
\begin{definition}[Programs as morphisms]
  \label{d:1:catsynt}
  In the following, we express a programming language by a category $L$.
  Its objects represent types/contexts of programs,
  and morphisms stand for programs.
\end{definition}
This approach has been often used to describe functional programs
(see e.g. \cite{ls:cup88,crole_1994}).
However, it also applies to other families of programs, such as imperative
languages, that may not seem, at first, ideally adapted to a categorical
approach.
We illustrate this in the following examples.
\begin{example}[A small imperative programming language]
  \label{e:1:while}
  We fix a set of values $\svals$ and a finite set of variables $\svars$.
  We define the category $\lwhile$ as follows.
  First, we let expressions be defined by the grammar $e \deq v \; (\text{where }
  v \in \svals) \mid x \; (\text{where } x \in \svars) \mid e \oplus e \;
  (\text{where }\oplus \in \{ +, -, \ast, \leq, =, \ldots \})$.
  Second, we let programs be defined by the grammar
  $P \deq \kwskip \mid P; P \mid x := e \mid \kwif \, e \, \{ P \} \kwelse
  \{ P \} \mid \kwwhile \, e \, \{ P \}$.
  Intuitively, an expression $e$ is a value, the reading of a variable, or
  a binary expression, and a program is either the $\kwskip$ program that
  does nothing, or a sequence, or an assignment, or a condition, or a
  loop.
  The sequence construction acts like a composition.
  To satisfy the properties of morphism composition, we need to equate
  $\kwskip; P$ and $P; \kwskip$ with $P$ (identity) and also to let
  sequential composition act as an associative operation (which boils down
  to not parenthesizing sequences).
  Therefore $\lwhile$ simply follows the typical single object category
  describing a monoid.
\end{example}
As the above example is rather contrived in the sense that it omits
any form of scoping, we consider a second, more sophisticated version.
\begin{example}[Imperative programs with scoped variables]
  \label{e:2:wscope}
  In this example, we extend \eref{1:while} with a notion of scope, based
  on explicit variable creation and destruction operations.
  We keep the definition of expressions unchanged and extend that of
  programs with two additional constructions $\kwaddvar \, x$ which
  adds a new local variable $x$ and stores a default value (that we
  assume to be $0$) into it and $\kwdelvar \, x$ which ends the scope
  of a local variable $x$.
  Note that conventional block-based scoping can be encoded using these
  two operations.

  We revise the construction of $\lwhile$ in \eref{1:while} and let
  objects be finite sets of variables.  Intuitively, the object $X$
  denotes set of memories storing $|X|$-many values.  To make this
  explicit, we stratify the set of programs by two sets of
  variables, that respectively denote the variables in their input and
  output states.
  More precisely, we specify the set $\lwhilev(X,X')$ of programs from
  $X$ to $X'$.
  Then, well-formed programs are defined inductively as follows.
  For any set $X$, $\kwskip \in \lwhilev(X,X)$.
  Given programs $P\in\lwhilev(X,X')$ and $P'\in\lwhilev(X',X'')$ the
  sequence program $P;P'$ is in the set $\lwhilev(X,X'')$.
  Assignment $x := e$ belongs to $\lwhilev(X,X)$ for any $X$ such that
  $x$ and all variables in $e$ are elements of $X$.
  Given two programs $P,P'\in\lwhilev(X,X')$, the condition program
  $\wcond e P {P'}$ is in $\lwhilev(X,X')$ when all variables in $e$ are
  in $X$.
  Similarly, loop $(\kwwhile \, e \, \{ P_{X,X} \})_{X,X}$ is defined when all
  variables in $e$ are elements of $X$ (note that the body and the loop
  program are morphisms from $X$ to itself).
  Finally, when $x \not\in X$,
  $(\kwaddvar \, x)\in\lwhilev(X,X \suplus \{x\})$ and
  $(\kwdelvar \, x)\in\lwhilev(X \suplus \{x\},X)$.
  We note that the sequence case acts as composition of two morphisms
  respectively from $X$ to $X'$ and from $X'$ to $X''$ and that it produces
  a morphism from $X$ to $X''$.
  The resulting category is denoted by $\lwhilev$.
\end{example}

\subsection{Semantic Interpretation}
Based on the category of programs setup in \sref{2:1:progs}, we now turn
to semantic interpretations.
Intuitively, a semantic interpretation should map programs into mathematical
objects that describe their behavior in a more abstract manner than just
syntax.
For instance, the interpretation of a program $P$ may boil down to a function
that maps program input states to corresponding output states.
Such an interpretation is expected to meet some compositionality property.
Therefore, functors appear as the right categorical tool to define semantic
interpretations.
Before we spell out any definition, we consider a couple of examples.

\begin{example}[Denotational Semantics]
  \label{e:3:semden}
  In this first example, we consider the language of \eref{1:while}.
  Programs are naturally regarded as actions on memories, namely functions from
  variables to values.  We thus first define $\smems \deq \fsp\svars\svals$ to
  be the set of memories. Since
  any memory state is a valid input/output for any program, the
  construction of \dref{1:catsynt} requires a single object that
  stands for $\smems$. This action will be given as the semantics in
  the next Example.  We observe that not all programs terminate, which
  entails that one input state may not correspond to exactly one
  output state.  Therefore, following the classical approach to
  denotational semantics, we interpret the unique object of $L$ into
  the set $\smemsbot \deq \smems \suplus \{ \dvg \}$ where $\dvg$
  stands for 
  non-termination.  Then, the interpretation of a program $P$ is the
  $\dvg$-strict function $\sem{P}$ from $\smemsbot$ into itself.  We
  remark that $\smemsbot$ is often interpreted as a poset, where
  $\forall m \in \smems, \dvg \sqsubseteq m$ and that $\dvg$-strict
  functions are monotone functions in that set.  Such interpretations
  compose well in the sense that the interpretation of $P_0; P_1$
  coincides with the composition of the interpretation of $P_1$ with
  that of $P_0$. Therefore we may convert it into a functor
  $\sem-:\lwhile\to\Pos$.  This is a {\em denotational semantics} of
  $\lwhile$, interpreting programs as {\em actions on states}
  (including $\dvg$). This type of semantics will be studied in
  \sref{densem}.

  This functorial semantics can be extended to the imperative language
  with variable scoping operators $\lwhilev$  in Example
  \ref{e:2:wscope}. In this language, by putting
  $\smems_X \deq \fsp X\svals$, program should be viewed a strict
  continuous function from $(\smems_X)_\dvg$ into $(\smems_{X'})_\dvg$
  for two given finite sets of variables $X, X'$.
  Note that, in this settings, $(\kwaddvar \, x)$ is well defined
  since it initializes the new variable $x$ with default value
  0, as specified in \eref{2:wscope}.
\end{example}
\begin{example}[Collecting semantics]
  \label{e:4:semcol}
  Another type of semantics is to interpret programs as actions on
  {\em properties} on states. There are various ways to represent
  them, and one way is by sets of states. Then the semantics of
  programs are expressed as functions from sets of states into sets of
  states.
  Such a semantics is usually called \emph{collecting semantics} and
  is one of the fundamental semantics in abstract interpretation.
  Specifically, we interpret the unique object of
  $\lwhile$ with the poset $(\partsof{\smems},\subseteq)$ and a
  program $P$ with a $\cup$-preserving function
  $\csem{P}: (\partsof{\smems},\subseteq) \rightarrow
  (\partsof{\smems},\subseteq)$.  Typically, $\csem{P}$ is defined by
  induction over the syntax of programs and is compositional in the
  same sense as $\sem{P}$ (\eref{3:semden}). It thus determines a
  functor $\csem-:L_w\to\Pos$.  It can also be derived rather
  systematically from $\sem{P}$ - see \sref{puttg}.
\end{example}
\begin{example}[Interval analysis]
  \label{e:5:ai:int}
  While the semantics shown in \eref{4:semcol} is suitable as a starting
  point to study static analyzers, it is not computable.
  Thus, we propose to consider a second interpretation, using abstract
  properties 
  rather than sets of 
  states~\cite{cc:popl:77}.
  We consider here the interval abstraction of~\cite{cc:popl:77}, assuming
  values are machine integers.
  In this setup a set of states is described either with $\bot$ (which
  denotes the empty set) or with a function from variables into intervals
  with integer bounds (that could be the minimum or maximum representable
  machine integers).
  Therefore, an interpretation of a program $P$ may be defined as an abstract
  semantics $\asem{P}$ that maps an abstract pre-condition into a sound
  abstract post-condition:
  given any abstract state $\amem$ and any memory $m$ that satisfies the
  constraints in $\amem$, all states in $\sem{P}(\{m\})$ are described by
  $\asem{P}(\amem)$.

  Nevertheless, this construction may not be compositional in the strict
  sense of denotational semantics.
  Indeed, the analysis of a composite program may be (and often is) more
  precise than the composition of the analyses of the sub-programs, in
  the following sense: we say that an analysis $\asem{P_0}$ is more
  precise than an analysis $\asem{P_1}$ when $\asem{P_0}$ computes
  properties that are logically stronger than those computed than
  $\asem{P_1}$.
  Indeed, let us consider $P_0$ be $x := 4 \ast x - 2$ and $P_1$ be
  $\kwif \, x \leq 0 \, \{ x = -x \} \kwelse \{ \kwskip \}$.
  Moreover, we let $\amem$ map $x$ to interval $[0,1]$.
  Then, $\asem{P_0}(x) = [-2, 2]$ and $\asem{P_1} \circ \asem{P_0}(x) =
  [0,2]$.
  However, if we consider concrete executions starting with $x \in [0,1]$,
  only value $2$ may be observed as an output, thus a more precise analysis
  $\asem{P_0; P_1}$ that maps $\amem$ to $[x \mapsto [2,2]]$ is sound.
\end{example}
The first remark that follows from the last two examples is that $\Pos$
provides a natural setup for property-based semantics.
Indeed, objects of $\Pos$ allow to account for collections of program
inputs or outputs, ordered with inclusion, with the usual meaning that
smaller elements account for fewer behaviors.

The second remark is that composition may not always be exact when
considering static analyses.
Indeed, while the concrete semantics of \eref{3:semden} is compositional,
\eref{5:ai:int} shows that static analyses may not be.
Though, common practice ensures that most analyses satisfy a lax form of
compositionality: the analysis of the composition of two programs may be
implemented so that it gives somewhat more precise results than that of
the composition of the analyses but is not expected to produce worse
results.
We comment on the limitations of this view in Remark~\ref{r:limit:oplax}.

To formalize these remarks, we rely on the order-enrichment
on the category $\Pos$ of posets and monotone functions.
All monotone functions $f,g:X\to Y$ in $\Pos$ are ordered in a pointwise
manner,
making $\Pos$ an enriched category over $\Pos$ itself.
Intuitively, the order $f \leq g$ means that ``$g$ is
\emph{more abstract} than $f$''/``$g$ is \emph{coarser} than $f$''.
Based on this terminology, we define semantic interpretations:
\begin{definition}[Interpretations]\label{d:intp}
  A \emph{semantic interpretation} (or, for short,
  \emph{interpretation}) $F$ consists of an object mapping
  $F: \Obj(L) \rightarrow \Obj(\Pos)$ and a morphism mapping
  $F_{X, Y} : L(X, Y) \rightarrow \Pos(F X, F Y)$ such that the
  following inequalities hold:
  \[
    F (\tmop{id}) \leq \tmop{id},
    \quad
    F (f_1 \circ f_2) \leq F f_1 \circ F f_2 .
  \]
  We say that the interpretation is \emph{normal} if
  $F (\tmop{id}_X) = \tmop{id}_X$ and \emph{functorial} if it is
  normal and $F (f_1 \circ f_2) = F f_1 \circ F f_2$.
\end{definition}
Such a weakened form of functor is known as \emph{oplax functors}. The
above definition makes sense when $\Pos$ is replaced with some other
$\Pos$-enriched category $K$. In fact, Steffen, Jay and Mendler
considered this general definition of interpretation in
\cite{sjm:tia:92}, and their intention is to pick a suitable $K$ for
each analysis task.
On the other hand, in this paper we fix the codomain of interpretation
to $\Pos$.

The semantics studied in \eref{3:semden} defines a functorial interpretation.
The analysis of \eref{5:ai:int} defines an interpretation that is
normal but not functorial.
This latter situation was also commented by Steffen, Jay and Mendler in
\cite{sjm:tia:92}:
indeed, they also introduce an oplax interpretation and remark that
strictness analysis is generally not functorial.
\begin{remark}[Limitations]
  \label{r:limit:oplax}
  As stated above, not all semantics satisfy the property of \dref{intp},
  and in particular, some static analyses fail to satisfy it.
  As an example, it is possible to set up a weaker form of interval analysis
  (\eref{5:ai:int}) that gives up all precision when applied to programs
  with updates to more than $k$ variables.
  If $k = 1$, $P_0$ is $x = 0$ and $P_1$ is $y = 1$, although these programs
  are trivial, $\asem{P_0; P_1}$ drops all information on $x, y$ whereas
  $\asem{P_1} \circ \asem{P_0}$ is trivially analyzed in a precise manner,
  thus we do not have $\asem{P_0 \circ P_1} \leq \asem{P_1} \circ
  \asem{P_0}$.
  Obviously, this definition is very contrived, and a conventional
  implementation of interval analysis achieves the oplaxness property
  stated in \dref{intp}.
\end{remark}

We introduce a partial order between interpretations. This order
$F\le G$ means that $F,G$ agree on the interpretation of objects in
$L$, but $F$ interprets programs with better precision than $G$.
\begin{definition}[Partial Order between Interpretations]\label{d:funccomp}
  Given two interpretations $F,G:L\to\Pos$, we write $F\le G$ if
  $F(a)=G(a)$ for any $a\in L$ and $F(f)\le G(f)$ for any $f:a\to b$
  in $L$.
\end{definition}
This order will be used to compare abstract interpretations derived by
Galois connections and its over-approximations
(Theorem \ref{t:galcon},\ref{t:absind},\ref{t:absindlam}).

\section{\GCS{} \confi{50}}
\label{s:3:colsem}
We have set-up a formulation of abstract semantics as oplax
functors. Typically, the development of abstract semantics is
initiated from a concrete semantics called {\em collecting semantics}.
We therefore express collecting semantics in our categorical setting.

The spirit of the collecting semantics is to interpret the behavior of
programs as actions on collections of inputs, rather than single
inputs. Typically, such a collection is chosen to represent a {\em
  property} on inputs, hence below we use the word {\em property} to
mean a collection of inputs, or a collection of elements in a set $X$
in general. Properties on $X$ are ordered by the inclusion,
forming the poset $(\pow X,\subseteq)$ (where $\pow X$ denotes the
set of subsets of $X$).

Let us consider a simple example of collecting semantics.
We consider a set-theoretic denotational semantics of a deterministic
and terminating programming language $L$.
It is defined by a functor $\sem-:L\to\Set$ mapping a
morphism $P:a\to b$ in $L$ representing a program into a function
$\sem P:\sem a\to \sem b$. Then, the {\em collecting semantics}
associated to this denotational semantics is the monotone function
$\csem{P}:(\pow{\sem a},\subseteq)\to(\pow{\sem b},\subseteq)$ defined
by $\csem{P}(U)\deq\{\sem{P}(x)~|~x\in U\}$; the right hand side
is the {\em direct image} of a property $U\subseteq\sem a$ by
$\sem P$. 
We then regard the collecting semantics as a functor $\csem-:L\to\Pos$.

We categorically analyze this definition of collecting semantics as
follows. We introduce the {\em covariant powerset functor}
$Q:\Set\to\Pos$ defined by:
\funcdefi{Q(X)}{(\pow{X},\subseteq)}{Q(f)(U)}{\{f(x)~|~x\in U\}.}  The
role of this functor is to assign to each $X\in\Set$ the poset of
properties on $X$, and to each morphism $f$ in $\Set $ the {\em direct
  image operation} associated to $f$.  Then we notice the equality
$\csem a=Q(\sem a)$ for each object $a\in L$ and $\csem P=Q(\sem P)$
for each morphism $P$ in $L$, which amounts to the
following functor equality:
\begin{displaymath}
  \csem- \quad = \quad
  \xymatrix@C=2cm{
    L \ar[r]^-{\sem-} & \Set \ar[r]^-{Q} & \Pos
  }.
\end{displaymath}

This functorial presentation of the collecting semantics suggests us
to generalize the middle category to an arbitrary category $\mcC$
instead of $\Set$ so that we can define the concept of collecting
semantics for general categorical semantics of programming
languages. However, we need to replace the covariant powerset functor
$Q:\Set\to\Pos$ with something else because it is specific to $\Set$. We therefore need to
supply a functor $C:\mcC\to\Pos$ that plays the same role as $Q$; it
assigns to each object $X\in\mcC$ a poset $CX$ of properties on $X$,
whose elements abstractly represent properties on $X$, and to each
$f:X\to Y$ a monotone function $Cf:CX\to CY$ representing the direct
image operation. Based on this observation, we derive the following
definition of collecting semantics.
\begin{definition}\confi{75}
  A {\em \gcs{}} of a language $L$ (regarded as a
  category) consists of a category $\mcC$ and two functors:
  \begin{displaymath}
    \xymatrix@C=2cm{
      L \ar[r]^-{\sem-} & \mcC \ar[r]^-{C} & \Pos
    },
  \end{displaymath}
  where $\sem -$ is called a {\em denotational semantics} and $C$ is
  called a {\em \propfun} (for $\mcC$). The composite
  $C\circ\sem -:L\to\Pos$ itself is also called the \gcs{}
  (with respect to $\sem-$).
\end{definition}
This definition of \gcs{} reflects our view that concrete semantics
for initiating abstract interpretation is {\em synthesized} from a
denotational semantics $\sem{-}:L\to\mcC$ by composing a {\em property
  functor} $C:\mcC\to \Pos$. However, in some situations one
may directly construct a collecting semantics without denotational
semantics ---in this case we simply put $C=\mathrm{Id}$.

\begin{remark}
  The readers who are familiar with fibrational category theory
  might notice that the property functor $C:\mcC\to\Pos$ bijectively
  corresponds to a {\em posetal opfibration} (a functor
  $c:\mcE\to\mcC$ from some category $\mcE$ such that $c$ satisfies
  the {\em opcartesian lifting property}; see e.g. \cite[Section
  9.1]{b:jacobs:00}). Therefore $C$ may be replaced
  by a posetal opfibration $c:\mcE \to\mcC$. In this setting, the
  \gcs{} interprets a morphism $f:a\to b$ in $L$ as the
  {\em pushforward} $\sem f_*:\mcE_{\sem a}\to\mcE_{\sem b}$ between fibre posets.
\end{remark}

\subsection{Denotational Semantics}\label{s:densem}
In this section, we focus on the first functor and look at several
examples of denotational semantics.
\begin{example}
  An extreme example of denotational semantics is the identity
  functor. This means that the semantics of types/contexts and
  programs are themselves.
\end{example}

A second non-trivial family of examples relates to \emph{monadic
  semantics}.
Good references to the definitions of {\em monads} and
{\em Kleisli categories} can be found in \cite{m:spr98}
and \cite{m:ic:91}.
The Kleisli category of a monad $T$ on $\mcC$ is denoted by $\mcC_T$.
We let $\bullet$ denote the composition of morphisms in Kleisli categories.
Among examples of monads on $\Set$, we can cite 1) the
\emph{maybe monad} that joins an extra element to a given set, and is
defined by $\mmay X=X\uplus\{*\}$, 2) the powerset monad which maps
each set to its powerset and is defined by $\mpow X=\pow X$, and 3)
the monad of probability subdistributions defined by
$\msdist X=\{\mu:X\to[0,1]~|~\text{$\mu$ is a countable subdist. on
  $X$}\}$.  Before we look at monadic semantics, we fix a class of
monads that is required to enable a least fixpoint definition of the
semantics of while commands.
\begin{definition}
  A {\em while-monad} on $\Set$ consists of a monad $(T,\eta,(-)^\#)$
  on $\Set$ and an $\omega$-cpo $(\sleq_X,\dvg_X)$ on each
  $TX$. The sup of an $\omega$-chain $\{x_i\}_{i\in\Nat}$ in $TX$ is denoted by $\bigsqcup x_i$. \footnote{ In this article, the least element of a poset is
    denoted by $\dvg_X$ if it corresponds to non-termination, while it
    is denoted by $\bot$ if it corresponds to the falsity
    (\eref{pfkleisli}).} We define $\sleq_{X,Y}$ as the pointwise
  order on $\Set_T(X,Y)$ given by
  $f\sleq_{X,Y}g \iff\fa{x\in X}f(x)\sleq_Y g(x)$, and $\dvg_{X,Y}$ as
  the least function $\lam{x}\dvg_Y$.  The data of the while-monad
  should satisfy:
  \begin{compactenum}
  \item the Kleisli composition $\bullet$ is monotone and
    $\omega$-continuous in each argument with respect to
    $\sleq_{X,Y}$, and
  \item additionally, it is strict in the second argument:
    $f\bullet \dvg_{X,Y}=\dvg_{X,Z}$ for any $f\in\Set_T(Y,Z)$.
  \end{compactenum}
\end{definition}
This is a simplification and specialization of the {\em order-enriched
  monad} in \cite{DBLP:conf/lics/GoncharovS13}. The order
$\sqsubseteq$ is to compare the definedness of elements/functions in
the sense of domain theory. We later see another order for truth
values, and these two orders are independent in general.  We do not
require $\bullet$ to be strict in the first argument.  This is because
we may sometimes want to distinguish the divergence after some effect
from pure divergence. Consider a program
$\mathtt{tick};\mathtt{diverge}$. Then its monadic semantics will be
$\dvg_{\smems,\smems}\bullet\sem{\mathtt{tick}}$, which we may want to
distinguish from the silent divergence $\dvg_{\smems,\smems}$.

Examples of while-monads on $\Set$ include the powerset monad $\mpow$
with the set inclusion order, the maybe monad $\mmay$ with the flat
order making $\iota_2(*)$ the least element, and the countable
subdistribution monad $\msdist$ with the pointwise mass order.
  
We now give a generic set-theoretic monadic semantics of the while
language in \eref{1:while}.
\begin{example}[Monadic Semantics of While Language]\label{e:monadwhile}
  We give a semantics of the while language in $\Set_T$.  Let
  $(T,\eta,(-)^\#,\sleq,\dvg)$ be a while-monad on $\Set$. In this
  semantics, we allow assignment commands $x:=e$ to perform some
  computational effects.
  We therefore let the interpretation of the command $x:=e$ be the
  morphism $\sem{x:=e}\in\Set_T(\smems,\smems)$.  We also assume
  an interpretation of the boolean expression $b$ as a function
  $\sem b\in\Set(\smems,\{\fff,\ttt\})$ into the two-points set.
  Then, the interpretation of programs is given by
  \begin{align*}
    \semt \kwskip
    \deq \eta_\smems \qquad
    \semt{P;P'}
    &\deq \semt{P'}\bullet\semt{P} \qquad
    \semt{\wasgn x e}
    \deq \sem{x:=e} \\
    \semt{\wcond b{P_1}{P_2}}
    &\deq \lam\rho\text{if $\sem b(\rho)=\ttt$ then $\semt{P_1}(\rho)$
      else $\semt{P_2}(\rho)$} \\
    \semt{\wwhile e P}
    &\deq  \mu\Phi\quad \text{where}\quad
    \Phi(f)
              \deq \lam\rho\text{if $\sem b(\rho)=\ttt$ then $f\bullet \semt P(\rho)$ else $\eta_{\powo{\smems}}(\rho)$}
  \end{align*}
  Here $\mu\Phi$ is the least fixpoint of $\Phi$, and $\eta_{\powo{\smems}}$
  is a component of the unit natural transformation of the monad $T$.
  We regard this interpretation as a functor
  $\semt-:\lwhile\to\Set_T$. When $T$ is the maybe monad $\mmay$,
  $\semt[\mmay]-$ interprets programs as partial functions, while when
  $T$ is the powerset monad $\mpow$, $\semt[\mpow]-$ may be regarded as
  interpreting programs as binary relations.
\end{example}
\begin{example}
  To interpret the while language $\lwhilev$ with variable addition
  and deletion (\eref{2:wscope}), we may adjust the semantics in the
  previous example as follows. First, we let $\semt-$ map each
  object $X\in\lwhilev$, which is a finite set of variables, to the
  set $\semt X\deq \smems_X$ of memories over $X$ (Example
  \ref{e:3:semden}).
  It is then almost straightforward to modify $\semt-$ and let it map
  a program $P\in \lwhilev(X,X')$ to a morphism of
  type $\smems_X\to\smems_{X'}$ in $\Set_T$. We interpret two
  additional commands $\kwaddvar$ and $\kwdelvar$ by
  \[
  \semt{(\kwaddvar~x)_{X,X\uplus\{x\}}}(\rho)
  = \eta_{\sem X\uplus\{x\}}(\rho\{x\mapsto 0\})
  \qquad
  \semt{(\kwdelvar~x)_{X\uplus\{x\},X}}(\rho)
  = \eta_{\sem X}(\rho-x),
  \]
  where $0$ is the presupposed default value for new variables
  (Example \ref{e:2:wscope}) and $\rho-x$ is the environment obtained
  by removing $x$ from the domain of $\rho$.
  Overall, we obtain a revised functor $\semt-:\lwhilev\to\Set_T$.
\end{example}

\subsection{\PropFun{}}

We next see some \propfuns{} so that we can form \gcs{} of
denotational semantics in the previous section.
\begin{example}
  An extreme example of a \propfun{} is the identity functor on
  $\Pos$. This assigns to each poset $X$ the poset of properties
  consisting of ``being less than or equal to $x$'' for each $x\in X$.
\end{example}
\begin{example}\label{e:pfrel}\confi{70}
  The category $\Rel$ of sets and binary relations between them is the
  host category for various relational semantics of programs. For
  instance, the relational semantics of the while language interprets
  a program as an endorelation on $\smems$.  Recall that each binary
  relation $f\in\Rel(X,Y)$ determines a $\cup$-preserving function
  $f_S(U)\deq \{j~|~i\in U\wedge (i,j)\in
  f\}:(\pow{X},\subseteq)\to(\pow{Y},\subseteq)$.  We turn this into
  a functor $I^\mpow:\Rel\to\CLat_\vee$, where $\CLat_\vee$ is the
  subcategory of $\Pos$ consisting of complete lattices and
  join-preserving functions between them, by \funcdefi
  {I^\mpow(X)}{(\pow{X},\subseteq)} {I^\mpow(f)}{f_S.}  Clearly we
  have the subcategory inclusion $\iota:\CLat_\vee\to\Pos$, so overall
  we obtain a \propfun{} for $\Rel$ as the composite
  $\iota \circ I^\mpow: \Rel \rightarrow \CLat_\vee \rightarrow \Pos$.
\end{example}
\begin{example}\label{e:pfkleisli}
  Recall that $\Rel$ is isomorphic to the Kleisli category
  $\Set_\mpow$ of the powerset monad $\mpow$. Thus the property
  functor in the previous example may be seen as the composite
  $\iota \circ I^\mpow: \Set_\mpow\rightarrow \CLat_\vee
  \rightarrow \Pos$. In this example, we generalize this diagram in
  two directions: one is to replace $\mpow$ with a $\Set$-monad
  $(T,\eta,(-)^\#)$, and the other is to adopt complete-lattice
  valued predicates as properties. We then
  derive a \propfun{} using the {\em strongest
    postcondition predicate transformer} for Kleisli categories
  studied in \cite{DBLP:conf/mfps/AguirreK20}.

  We first take a complete lattice $(\Omega,\le)$ for truth values.
  We use the symbols $\bot,\top,\vee,\wedge$ to mean the least/greatest
  elements and joins/meets of $\Omega$. We then regard a
  function $\phi:X\to\Omega$ as an {\em $\Omega$-valued property on
    $X$}.  Such properties form a complete lattice
  $\powo X\deq (\Set(X,\Omega),\le_X)$, where $\le_X$ is the
  pointwise order. The lattice operations on $\powo X$ is
  written by $\wedge_X,\vee_X$, etc. We note that the mapping
  $X\mapsto \powo X$ extends to a functor of type $\Set^{op}\to\CLat$
  (the category of complete lattices and complete lattice homomorphisms).

  We next take a $T$-{\em algebra} (or {\em Eilenberg-Moore $T$-algebra})
  $o:T\Omega\to\Omega$ (see \cite[Section VI.2]{m:spr98}), and assume
  that $o$ is {\em meet-preserving} in the following sense: the
  function $\lam{\phi}o\circ T(\phi)$ belongs to
  $\CLat_\wedge(\powo X,\powo{T(X)})$, the homset of the category
  $\CLat_\wedge$ of complete lattices and meet-preserving
  functions.
  Such a $T$-algebra induces the {\em weakest precondition predicate transformer}
  $wp^o(f)\in \CLat_\wedge(\powo Y,\powo X)$ for each
  $f\in\Set_T(X,Y)$. It is given by
  $wp^o(f)\deq\lam\phi o\circ T\phi\circ f$ \cite[Corollary
  4.6]{DBLP:conf/mfps/AguirreK20}. The mapping $f\mapsto wp^o(f)$
  extends to a functor of type $\Set_T^{op}\to\CLat_\wedge$, thanks to
  the Eilenberg-Moore axioms.

  We then take the left adjoint $sp^o(f)\in\CLat_\vee(\powo X,\powo Y)$ of
  $wp^o(f)$, which we call the {\em strongest postcondition predicate
    transformer} \cite[Example 4.11]{DBLP:conf/mfps/AguirreK20}. We
  extend the mapping $f\mapsto sp^o(f)$ to a functor
  $I^{o}:\Set_T\to \CLat_\vee$ given by
  \funcdefi{I^{o}(X)}{\powo X}{I^{o}(f)}{sp^o(f)}. In this way we obtain a
  \propfun{}
  $\iota \circ I^o: \Set_T \rightarrow \CLat_\vee \rightarrow \Pos$.

  We see a few examples of meet-preserving $T$-algebras
  where $\Omega=\{\bot\le\top\}$. Now
  $\powo X$ is the poset of characteristic functions, and we identify
  it with the poset $(\pow X,\subseteq)$ of subsets of $X$.
  \begin{enumerate}
  \item For the powerset monad $\mpow$, there is only one
    $T$-algebra $o:\mpow\Omega\to\Omega$ preserving meets:
    $o(U)=\top\iff \bot\not\in U$ \cite[Example
    5.3]{DBLP:conf/mfps/AguirreK20}. The derived \propfun{} is
    isomorphic to the one in \eref{pfrel}.

  \item For the maybe monad $\mmay$, there is only one $T$-algebra
    $o:\mmay\Omega\to\Omega$ preserving meets: the one sending
    $\iota_2(\dvg)$ to $\top$. The strongest postcondition predicate
    transformer (hence \propfun{} $I^o$) satisfies, for
    $f\in\Set_\mmay(X,Y)$,
    $sp^o(f)=\lam\phi\{y\in Y~|~\ex{x\in X} x\in\phi\wedge
    f(x)=\eta_Y(y)\}$.
  \end{enumerate}
\end{example}
\begin{example}\label{e:pfforget}\label{e:pfkleisliforget}\confi{70}
  Any category $\mcC$ with a functor $U:\mcC\to\Set$ has a \propfun{}
  given by the composite
  $Q \circ U: \mcC \rightarrow \Set \rightarrow \Pos$ with the covariant
  powerset functor $Q$.

  For example, let $(T,\eta,(-)^\#)$ be a monad on $\Set$. Its Kleisli
  category $\Set_T$ comes with the right adjoint functor
  $K^T:\Set_T\to\Set$ given by \funcdefi {K^T(X)}{T(X)}{K^T(f)}{f^\#}
  \cite[Theorem VI.1]{m:spr98}. We then compose this with the
  covariant powerset functor $Q$ and obtain a \propfun{}
  $\cpow \circ K^T: \Set_T \rightarrow \Set \rightarrow \Pos$.

  By letting $T$ be the powerset monad $\mpow$, we obtain the
  \propfun{} that assigns to a set $X$ the set of {\em hyperproperties}
  over $X$ in the sense of Clarkson and Schneider \cite{cs:jcs:10}.
  The object part of $Q\circ K^T$ sends a set $X$ to the
  poset $(\pow{\pow{X}},\subseteq)$ of hyperproperties on $X$.
\end{example}

\subsection{Putting Together \confi{60}}\label{s:puttg}

We have seen several denotational semantics and \propfuns{}.
By combining them we obtain \gcs{}.

We consider a \gcs{} of the while language $\lwhile$ with 1) the
monadic denotational semantics $\sem-_T:\lwhile\to \Set_T$ in
\eref{monadwhile}, and 2) the \propfun{}
$\iota\circ I^o:\Set_T\to\Pos$ given by the strongest postcondition
predicate transformer $sp^o$ in \eref{pfkleisli}.  The \gcs{}
interprets a program $P$ as
$\csem P\deq \iota\circ I^o(\sem P_T)=sp^o(\sem P_T)$.

We address the question of whether this collecting semantics can be
given inductively. Let $(T,\eta,(-)^\#,\sleq,\dvg)$ be the while-monad
on $\Set$ used in the monadic denotational semantics, $(\Omega,\le)$
be the complete lattice, and $o:T\Omega\to\Omega$ be the
meet-preserving $T$-algebra used in the \propfun{}. We remark that
these two orders represent two independent notions: $\sleq$ represents
the definedness order of computations in the sense of domain theory,
while $\le$ represents the strength of truth values in the sense of
algebraic logic.
As noted in \cite{cc:jlc:92}, we emphasize these two orders may be
different in general.
We also use different symbols for these two orders:
joins for a definedness order is denoted by $\sqcup$.

The following theorem states that the collecting semantics
of while-free programs can be inductively given.
\begin{theorem}\label{t:induct}
  The \gcs{} $\csem -$ satisfies:
  \begin{align*}
    \csem \kwskip
    = \id_{\powo \smems} \qquad
    \csem{P;P'}
    &= \csem{P'}\circ\csem{P} \qquad
    \csem{\wasgn x e}
    = \textstyle sp^o(\sem{x:=e}) \\
    \csem{\wcond b{P_1}{P_2}}
    &= \lam \phi \csem{P_1}(\phi\wedge_{\smems} \grd b \ttt)\vee_{\smems} \csem{P_2}(\phi\wedge_{\smems} \grd b {\fff}).
  \end{align*}
  Here,
  $\grd b v:\smems\to \{\bot_\smems,\top_\smems\}\subseteq \powo\smems$
  is the function defined by $\grd b v(\rho)=\top_\smems$ if and only
  if $\sem b(\rho)=v$.
\end{theorem}

The next question is whether
$\csem{\wwhile b P}\in\CLat_\vee(\powo{\smems},\powo{\smems})$ can be
computed by the least fixpoint of some functional, say $\Psi$,
definable by $P$ and $b$. Since $\semt{\wwhile b P}$ is the least
fixpoint of a functional $\Phi$ on $\Set_T(\smems, \smems)$ (see
\eref{monadwhile}), it suffices to show that 1)
$sp^o:\Set_T(\smems,
\smems)\to\CLat_\vee(\powo{\smems},\powo{\smems})$ is
$\omega$-continuous and strict, and 2)
$\Psi\circ sp^o=sp^o\circ \Phi$. Regarding 1, it is equivalent to the
continuity of $o:T\Omega\to\Omega$ in the following sense. It relates
the $\omega$-lub in the definedness order and the $\omega$-lub in the
truth value order.
\begin{proposition}\label{p:sp}
  The following hold:
  \begin{enumerate}
  \item $sp^o(\bigsqcup f_i)=\bigvee sp^o(f_i)$ holds for
     any $X,Y\in\Set_T$ and 
    $\omega$-chain $f_i$ in hom-$\omega$-cpo
    $(\Set_T(X,Y),\sqsubseteq_{X,Y})$ if and only if
    $o(\bigsqcup c_i)=\bigwedge (o(c_i))$ holds for any $\omega$-chain
    $c_i$ in the $\omega$-CPO $(T\Omega,\sleq_\Omega)$.
    \label{p:sp-meet}
  \item
    $sp^o(\dvg_{X,Y})=\lam{\phi}\bot_Y$
    if and only if $o(\dvg_{\Omega})=\top$.
    \label{p:sp-bot}
  \end{enumerate}
\end{proposition}
\begin{proof}
  We here prove (i); (ii) can be proved similarly.
  We actually prove that the following are equivalent.
  \begin{enumerate}[label=(\alph{enumi})]
  \item $sp^o\left(\bigsqcup f_i\right)=\bigvee sp^o(f_i)$ for any $X,Y\in\Set_T$ and $\omega$-chain
    $f_i$ in hom-$\omega$-CPO $(\Set_T(X,Y),\sleq_{X,Y})$. \label{en:sp}
  \item $wp^o\left(\bigsqcup f_i\right)=\bigwedge wp^o(f_i)$ for any $X,Y\in\Set_T$ and  $\omega$-chain
    $f_i$ in hom-$\omega$-CPO $(\Set_T(X,Y),\sleq_{X,Y})$. \label{en:wp}
  \item $o\left(\bigsqcup c_i\right)=\bigwedge (o(c_i))$ for any $\omega$-chain
    $c_i$ in the $\omega$-CPO $(T\Omega,\sleq_\Omega)$. \label{en:alg}
  \end{enumerate}
  
  \ref{en:sp} $\iff$ \ref{en:wp} Let $f_i$ be an $\omega$-chain in
  $(\Set_T(X,Y),\sleq_{X,Y})$. Then from $sp^o(f_i)\dashv wp^o(f_i)$,
  we obtain:
  \begin{align*}
    sp^o\left(\bigsqcup f_i\right)=\bigvee sp^o(f_i)
    \iff & \left(\fa{\phi,\psi}sp^o\left(\bigsqcup f_i\right)(\phi)\le_Y\psi\iff\bigvee sp^o(f_i)(\phi)\le_Y\psi\right)\\
    \iff & \left(\fa{\phi,\psi}\phi\le_X wp^o\left(\bigsqcup f_i\right)(\psi)\iff \phi\le_X \bigwedge wp^o(f_i)(\psi)\right)\\
    \iff & wp^o\left(\bigsqcup f_i\right)=\bigwedge wp^o(f_i).
  \end{align*}

  \ref{en:wp} $\implies$ \ref{en:alg} Let $c_i$ be an $\omega$-chain in
  $(T\Omega,\sleq_\Omega)$. Notice the isomorphism
  $(T\Omega,\sleq_\Omega)\cong
  (\Set_T(1,\Omega),\sleq_{1,\Omega})$. We thus identify $c_i$ as an
  $\omega$-chain in $(\Set_T(1,\Omega),\sleq_{1,\Omega})$. Then
  \begin{align*}
    & o\circ\left(\bigsqcup c_i\right)
      =wp^o\left(\bigsqcup c_i\right)(\id_\Omega)
      =\bigwedge wp^o(c_i)(\id_\Omega)
      =\bigwedge o\circ c_i.
  \end{align*}

  \ref{en:alg} $\implies$ \ref{en:wp} Let $f_i$ be an $\omega$-chain in
  $(\Set_T(X,Y),\sleq_{X,Y})$. Then for any $\phi:Y\to\Omega$ and
  $x\in X$, we have
  \begin{align*}
    & wp^o\left(\bigsqcup f_i\right)(\phi)(x)=
      o\circ T\phi\circ \left(\bigsqcup f_i\right)(x)=
      o\left(T\phi\left(\bigsqcup f_i(x)\right)\right)=
      o\left(\bigsqcup T\phi(f_i(x))\right)\\
    =&
       \bigwedge o(T\phi(f_i(x)))=
       \bigwedge (wp^o(f_i)(\phi)(x))=
       \left(\bigwedge wp^o(f_i)\right)(\phi)(x).
  \end{align*}
  Notice that $T\phi$ is continuous as $T$ is a while monad.
  Therefore $wp^o\left(\bigsqcup f_i\right)=\bigwedge wp^o(f_i)$.
\end{proof}

We thus say that $o$ {\em makes $sp^o$ $\omega$-continuous and strict}
if it is a strict $\omega$-continuous function of type
$(T\Omega,\sqsubseteq_\Omega)^{op}\to(\Omega,\le)$.
\begin{theorem}\label{t:inductwhile}
  If $o$ makes $sp^o$ $\omega$-continuous and strict, the \gcs{}
  $\csem -$ satisfies
  \begin{displaymath}
    \csem{\wwhile b P}=
    \mu\Psi
    \quad\text{where}\quad
    \Psi(f)=\lam \phi
    (f\circ \csem P(\phi\wedge_{\smems} \grd b \ttt))\vee_{\powo{\smems}}
    (\phi\wedge_{\smems} \grd b \fff).
  \end{displaymath}
\end{theorem}
\begin{proof}
  It sufficies to show $sp^o(\mu\Phi)=\mu\Psi$.  It is easy to verify
  that 
  $\Psi$
  is a join-preserving endofunction on the pointwise-order complete lattice
  $\CLat_\vee(\powo{\smems},\powo{\smems})$.  To
  show that $(-)\wedge_{\smems} \grd b v$ preserves joins, we use
  the fact that $\grd b v$ takes only values in $\{\bot_{\smems},\top_{\smems}\}$. By calculation, we have $\Psi\circ sp^o=sp^o\circ\Phi$.
  Since $sp^o$ is $\omega$-continuous and strict, we obtain
  $sp^o(\mu\Phi)=\mu\Psi$.
\end{proof}

We see some examples of meet-preserving
$T$-algebra making $sp^o$ $\omega$-continuous and strict.
\begin{enumerate}
\item For the powerset while-monad $\mpow$ on $\Set$ and
  $\Omega=\{\bot\le\top\}$, the meet-preserving $T$-algebra $o$
  in \eref{pfkleisli} makes $sp^o$ $\omega$-continuous and
  strict. The \gcs{} $\csem-$ coincides with the standard one in the
  literature.
\item For the powerset while-monad $\mpow$ on $\Set$ and
  $\Omega=([0,\infty],\le)$, the inf-operation
  $\inf:\mpow[0,\infty]\to [0,\infty]$ is a meet-preserving
  $T$-algebra making $sp^o$ $\omega$-continuous and
  strict. The \gcs{} $\csem-$ coincides with the {\em quantitative
    strongest postcondition} in \cite{10.1145/3527331}. By taking
  the opposite complete lattice $([0,\infty],\ge)$ for $\Omega$ and
  replacing $\inf$ with $\sup$, the \gcs{} $\csem-$ coincides with
  the {\em quantitative strongest liberal postcondition} in
  \cite{10.1145/3527331}.
\item For the maybe while-monad $\mmay$ and
  $\Omega=\{\bot\le\top\}$, the meet-preserving
  $T$-algebra $o$ in \eref{pfkleisli} makes $sp^o$
  $\omega$-continuous and strict.
\end{enumerate}

\begin{example}
  We consider a \gcs{} of the while language $\lwhile$ with the
  monadic denotational semantics $\semt-:\lwhile\to \Set_T$ in
  \eref{monadwhile} and the \propfun{} $Q\circ K^T:\Set_T\to\Pos$ in
  \eref{pfkleisliforget}. The interpretation
  $Q\circ K^T(\semt P):\pow{T\smems}\to\pow{T\smems}$ of a program $P$
  by this \gcs{} satisfies
  $ Q\circ K^T(\semt P)(U)=\{\semt P^\#(c)~|~c\in U\}, $ for any
  $U\in \pow{T\smems}$, where $(-)^\#$ is the Kleisli lifting of the
  monad $T$.  When $T$ is the powerset monad $\mpow$, this \gcs{}
  appears in Asaf et al.'s study on {\em hypercollecting semantics}
  \cite[Section 4]{10.1145/3009837.3009889}. The definition of their
  hypercollecting semantics is partially inductive\footnote{The
    hypercollecting semantics in \cite[Section
    4]{10.1145/3009837.3009889} is not inductive at the interpretation
    of conditional expressions.  }, and is an over-approximation of
  the above \gcs{}; see the proof of Theorem 1 of their
  paper.

\end{example}

\section{Semantic Abstraction}
\label{s:4:abs}
After discussing the role of collecting semantics in detail in
\sref{3:colsem}, we now consider semantic abstraction and derivation
of abstract semantics from a reference (e.g., collecting) semantics.

\subsection{Abstraction Relations Between Domains and Interpretations}
\label{s:absrel}
So far, we have studied the definition of program semantics independently
from one another and have not considered comparing them quite yet.
Abstract interpretation~\cite{cc:popl:77} is specifically motivated with
semantic comparison so as to tie properties that may be proved with one
to statements involving another.
Therefore, we consider the comparison of program semantics here.

Several forms of \emph{abstraction relations} have been proposed,
including Galois connections~\cite{cc:popl:77}, concretization functions
without assuming the existence of a best abstraction~\cite{cc:popl:79},
abstraction functions without assuming the existence of a concretization,
or binary relations~\cite{cc:jlc:92,cc:lp:92}.
In categorical terms, a {\em Galois connection} $\galois \alpha \gamma A C$
is a pair of monotone functions $\alpha:C\to A$ and $\gamma:A\to C$ between posets such
that the equivalence $\alpha(a)\le c\iff a\le\gamma(c)$ holds for any
$a\in A,c\in C$.
It is called a {\em Galois insertion} if $\alpha\circ\gamma=id$.

\begin{definition}\label{d:concabs}
  Let $A,C:L\to\Pos$ be \intps{}.

  A {\em \concd{}} from $A$ to $C$ is a family
  $\{\gamma_a:A(a)\to C(a)\}_{a\in L}$ of monotone functions. We say
  that it is a {\em \conci{}} from $A$ to $C$ if it satisfies
  $C(f)\circ \gamma_a\le \gamma_b\circ A(f)$ for any $f\in L(a,b)$.
  Dually, an {\em \absd{}} from $C$ to $A$ is a family
  $\{\alpha_a:C(a)\to A(a)\}_{a\in L}$ of monotone functions.  We say
  that it is an {\em \absi{}} from $C$ to $A$ if it satisfies
  $ \alpha_b\circ C(f)\le A(f)\circ \alpha_a $ for any $f\in L(a,b)$.
  A concretization (resp., abstraction) of interpretations is called
  {\em complete} if the above inequality is an equality.
\end{definition}
Remark that this definition sets up \emph{two} notions of concretization
(and the same for abstractions): the notion of \concd{} ties only semantic
domains, just like the concretizations in~\cite{cc:popl:77} do whereas
\conci{} tie not only domains (via a \concd{}) but also interpretations,
thus providing an interpretation soundness statement as typically sought
to design a static analysis.

In categorical terminology, a \conci{} is exactly a {\em lax natural
  transformation} from $A$ to $C$, and an \absi{} is exactly an {\em
  oplax natural transformation} from $C$ to $A$.\footnote {The
  original definition of (op)lax natural transformation consists of
  families of 1-cells and 2-cells. In the current context, however,
  there is at most one choice of 2-cells for (op)lax natural
  transformations between oplax functors into $\Pos$. Therefore we here treat the concept of (op)lax
  natural transformation as a property on families of 1-cells.  } Both
concretizations and abstractions of interpretation express that one
\intp{} is more abstract than the other.

Moreover, as in \cite{cc:popl:77}, we remark that the same abstraction
relation may be described simultaneously both by \concd{} and by an
\absd{}; this strong correspondence between the two then defines a
Galois connection.

We now explain how we develop a sound abstract interpretation of a
\gcs{} $(\sem-,C)$ in our categorical framework.
The ultimate goal is to construct an interpretation $A$ together with
a \conci{}, as shown in the diagram \eqref{eq:abs}.
The development process is broken into the following steps:
\begin{enumerate}
\item We give the object part of $A$; this corresponds to
  designing a domain of interpretation for each type/context
  $a\in L$.
\item We give a \concd{} $\{\gamma_a:A(a)\to C\sem a\}$, relating the
  domain of interpretation of types/contexts and that of the
  collecting semantics.
\item We give the morphism part of $A$; this corresponds
  to designing how each program $P:a\to b$ transfers abstract
  properties in $A(a)$ to those in $A(b)$.  In our categorical
  framework this should respect the oplax functoriality in
  \dref{intp}.
\item We check if the constructed interpretation $A$ satisfies
  soundness in the following sense: for any program $P:a\to b$ in $L$
  and an abstract predicate $\phi\in A(a)$, we have
  $ C(\sem P)(\gamma_a(\phi))\le \gamma_b(A(P)(\phi))$.  This is
  equivalent to showing that $\gamma$ being a \conci{} from $A$ to the
  \gcs{} $C\circ\sem-$.
\end{enumerate}
The delicate part of this process is to find a right combination of
$A(a)$, $\gamma_a$ and $A(P)$ to achieve the soundness, as well as the
expressiveness of abstract properties and the effectiveness of the semantics
(when implementing it on a computer).
In many cases, it is possible to simultaneously carry out the third and
fourth steps, and to derive $A$ from the interpretation $C\circ\sem-$ and the
abstraction relation, so as to achieve a sound $A$.


In the rest of the section we see examples of \conci{}. A bigger example
of the development of abstraction is given for the denotational
semantics of the lambda calculus (\sref{systemt}).
\commentout{
\begin{example}
  Let $\Aff$ be the category of affine functions: objects are
  natural numbers, and a morphism $f:n\to m$ in $\Aff$ is an affine
  function of type $\real^n\to\real^m$. We regard this
  as an elementary programming language.

  For each $n\in \Aff$, define $Ph(n)$ to be the set of convex
  polyhedra~\cite{ch:popl:78} in $\real^n$, ordered by inclusion.
  Each affine function $f:n\to m$ sends a polyhedron $P\in Ph(n)$
  into the {\em direct-image polyhedron} $f(P)=\{f(v)~|~v\in P\}\in
  Ph(m)$ thanks to Fourier-Motzkin theorem.
  This induces a functor $Ph:\Aff\to\Pos$.

  There is a denotational semantics functor $U:\Aff\to\Set$ given by
  $U(n)=\real^n$ and $U(f)=f$. Then the poset inclusion
  $\gamma^{Ph}_n:Ph(n)\subseteq\pow{\real^n}$ becomes a {\em natural
    transformation} $\gamma^{Ph}:Ph\to Q\circ U$, hence a
  \conci{}.
\end{example}
}

\subsection{The Case of Galois Connections}
\label{s:4:2:gcabsint}
As discussed earlier, Galois connections are one of the fundamental
constructions that are used in abstract interpretation to describe
abstraction relations.
In presence of such a strong connection, the semantic induction
technique of an \intp{} that was mentioned in the previous subsection
can be made even more systematic.
We now study it in our categorical framework.
\begin{theorem}
  \label{t:galcon}
  Let $C:L\to\Pos$ be an (resp. normal) \intp{},
  $\{A(a)\}_{a\in L}$ be a family of posets indexed by objects in $L$,
  and
  $G\deq\{\galois{\alpha_a}{\gamma_a}{A(a)}{C(a)}\}_{a\in L}$ be a
  family of Galois connections (resp. insertions). We define
  mappings on $L$-objects and $L$-morphisms by
  \funcdefi{C^G(a)}{A(a)}{C^G(f)}{\alpha_b\circ Cf\circ\gamma_a\quad (f:a\to b).}
  \begin{enumerate}
  \item $C^G:L\to\Pos$ is a (resp. normal) \intp{}, which we call the
    \intp{} {\em induced by $G$}.
  \item $\gamma$ is a \conci{} from $C^G$ to $C$. Moreover, for
    any \intp{} $A':L\to\Pos$ such that $A'(a)=A(a)$, if $\gamma$ is a
    \conci{} from $A'$ to $C$, then $C^G\le A'$.
  \item $\alpha$ is an \absi{} from $C$ to $C^G$. Moreover, for
    any \intp{} $A':L\to\Pos$ such that $A'(a)=A(a)$, if $\alpha$ is an
    \absi{} from $C$ to $A'$, then $C^G\le A'$.
  \end{enumerate}
\end{theorem}
\begin{proof}
  (i) $C^G(\id_a) \le\id_{Ca}$ is immediate by
  $\alpha_a\circ\gamma_a\le\id_{Ca}$. We show the other inequality:
  \begin{align*}
    C^G(g\circ f) &= \alpha_{c}\circ Cg \circ Cf\circ \gamma_{a}
                    \le \alpha_{c}\circ Cg\circ \gamma_{b}\circ\alpha_{b}\circ Cf\circ \gamma_{a} 
                    = C^G(g)\circ C^G(f).
  \end{align*}
  (ii) We show $\gamma$ is a lax natural transformation from $C^G$ to
  $C$, that is, $C(f)\circ \gamma_a\le \gamma_b\circ C^G(f)$. This is
  evident as
  $C(f)\circ \gamma_a\le \gamma_b\circ \alpha_b\circ
  C(f)\circ\gamma_a$.  Next, let $A':L\to\Pos$ be an interpretation
  such that $A'(a)=A(a)$ and assume that $\gamma_a$ is a
  concretization of interpretation from $A'$ to $\Pos$. This amounts
  to $C(f)\circ \gamma_a\le \gamma_b\circ A'(f)$ for any $f:a\to b$ in
  $L$.  Since $\alpha_b$ is a left adjoint of $\gamma_b$, this is
  equivalent to $\alpha_b\circ C(f)\circ \gamma_a\le A'(f)$, that is,
  $C^G(f)\le A'(f)$. (iii) can be proved similarly.
\end{proof}

When considering a program $f:a\to b$, the interpretation
$\alpha_b\circ Cf\circ\gamma_a$ is often called \emph{best abstract
  transformer}.
The above theorem says that $C^G$ is the most precise \intp{}
making $\gamma$ a \conci{} from $C^G$ to $C$, as well as
$\alpha$ an \absi{} from $C$ to $C^G$.

The induction of abstract semantics by Galois connections is a
fundamental operation in abstract interpretation, and our categorical
framework of abstract interpretation accommodates it by the above
theorem. Employing oplax functors as interpretations (\dref{intp}) is
crucial here, because ordinary functors are not closed under the
induction operation by Galois connection.

\paragraph{An abstraction of an interpretation in
  a complete lattice defining a Galois connection.}
The semantics induced by a family of Galois connections would not have
an inductive characterization due to its oplaxness.  The common
practice is to derive the best abstract transformers (or approximate ones)
for the basic constructs of a language, then extend them to the whole
language by induction.  We discuss this approach using the \gcs{} of
the while language in \sref{puttg} using a meet-preserving
Eilenberg-Moore $T$-algebra $o:T\Omega\to\Omega$ and a Galois
connection $G\deq \galois \alpha \gamma A {(\powo{\smems},\le)}$ with
a general complete lattice $(A,\wedge_A,\vee_A)$; later we restrict
$A$ so that the abstract interpretation can be computed in a finite
means. We apply this Galois connection to the \gcs{} $\csem-$ in
\sref{puttg}, and induce an abstract interpretation
$\csem -^G:\lwhile\to\Pos$, which is an oplax functor. It interprets a
program $P$ as a monotone function
$\alpha\circ\csem{P}\circ\gamma:A\to A$, but it does not enjoy an
inductive characterization, that is, $\csem{P}^G$ is not expressible
by the interpretation of subprograms in $P$ by $\csem{-}^G$. We
therefore inductively construct another interpretation $\asem-$ that uses
the best abstract transformers at assignment commands:
\begin{align*}
  \asem \kwskip
  \deq \id_{{A}} \qquad
  \asem{P;P'}
  & \deq \asem{P'}\circ\asem{P} \qquad
  \asem{\wasgn x e}
  \deq \csem{\wasgn x e}^G(=\alpha\circ sp^o(\sem{x:=e})\circ\gamma) \\
  \asem{\wcond b{P_1}{P_2}}
  & \deq \lam \phi \asem{P_1}(\phi\wedge_{A} \alpha(\grd b \ttt))\vee_{A}
  \asem{P_2}(\phi\wedge_{A} \alpha(\grd b {\fff})) \\
  \asem{\wwhile b P}
    & \deq \mu\Theta
      \quad\text{where}\quad
      \Theta(f)\deq\lam \phi
      (f\circ \asem P(\phi\wedge_{A} \alpha(\grd b \ttt)))\vee_{A}
      (\phi\wedge_{A} \alpha(\grd b \fff))
\end{align*}
Recall that $sp^o(\sem{x:=e})$ denotes the strongest postcondition
predicate transformer for assignments (\eref{pfkleisli}), and
$\grd b v$ denotes the predicate representing the condition tests
(\tref{induct}).  In the interpretation of $\kwwhile$, we use the
complete lattice structure on the homset $\Pos(A,A)$ with the
pointwise order.
\begin{theorem}\label{t:absind}
  Suppose that $o$ makes $sp$ $\omega$-continuous and strict (hence
  \tref{inductwhile} holds).  The interpretation $\asem-:\lwhile\to \Pos$
  is functorial and $\csem -^G\le \asem -$ holds.
\end{theorem}
\begin{proof}
  The functoriality of $\asem-$ is obvious by definition. We show
  $\alpha\circ\csem{P_0}\circ\gamma\le\asem{P_0}$ by induction on the
  structure of $P_0$.  We omit subscript of $\wedge,\vee$. The
  cases $P_0=\kwskip,(P;P'),(\wasgn x e)$ are easy.
  The case $P_0=\wcond b {P_1} {P_2}$ is shown as follows.
  \begin{align*}
    \asem{\wcond b {P_1} {P_2}}
    & = \lambda \phi . \asem{P_1} (\phi \wedge \alpha (\grd b \ttt)) \vee \asem{P_2} (\phi \wedge \alpha (\grd b \fff))\\
    \nt{IH}
    & \ge \lambda \phi . \alpha \circ \csem{ P_1 }
      \circ \gamma (\phi \wedge \alpha (\grd b \ttt)) \vee \alpha
      \circ \csem{ P_2 } \circ \gamma (\phi \wedge \alpha
      (\grd b \fff))\\
    \nt{$\gamma$ meet-pres.}
    & = \lambda \phi . \alpha \circ \csem{ P_1 } (\gamma
      (\phi) \wedge \gamma( \alpha (\grd b \ttt))) \vee \alpha \circ
      \csem{ P_2 } (\gamma (\phi) \wedge \gamma (\alpha (\grd b \fff)))\\
    \nt{unit law and $\alpha$ join-pres.}
    & \ge \lambda \phi . \alpha (\csem{ P_1 } (\gamma
      (\phi) \wedge \grd b \ttt) \vee \csem{ P_2 }
      (\gamma (\phi) \wedge \grd b \fff))\\
    \nt{definition}
    & = \alpha \circ \csem{\wcond b{P_1}{P_2}} \circ \gamma
  \end{align*}
  To show the case $P_0=\wwhile b P$, we compare $\Psi$ defined in
  \tref{inductwhile} and $\Theta$ used in the definition of $\asem{\wwhile b P}$.
  We first show $\alpha\circ\Psi(f)\circ\gamma\le\Theta(\alpha\circ f\circ\gamma)$.
  \begin{align*}
    \alpha \circ \Psi (f) \circ \gamma
    & = \lambda \phi . \alpha (f (\csem{ P } (\gamma (\phi) \wedge \grd b {\ttt})) \vee
      (\gamma (\phi) \wedge \grd b {\fff})) \\
    \nt{unit law and $\alpha$ join-pres.}
    & \le \lambda \phi . (\alpha \circ f \circ \csem{ P   } (\gamma (\phi) \wedge \gamma (\alpha (\grd b {\ttt})))) \vee \alpha (\gamma (\phi) \wedge \gamma (\alpha (\grd b {\fff}))) \\
    \nt{$\gamma$ meet-pres.}
    & = \lambda \phi . (\alpha \circ f \circ \csem{ P } \circ \gamma (\phi \wedge \alpha (\grd b {\ttt}))) \vee \alpha (\gamma (\phi \wedge \alpha (\grd b {\fff}))) \\
    \nt{unit and counit law}
    & \le \lambda \phi . (\alpha \circ f \circ \gamma \circ \alpha \circ \csem{ P } \circ \gamma (\phi \wedge \alpha (\grd b {\ttt}))) \vee (\phi \wedge \alpha (\grd b {\fff}))\\
    \nt{IH}
    & \le \lambda \phi . (\alpha \circ f \circ \gamma \circ \asem P (\phi \wedge \alpha (\grd b {\ttt}))) \vee (\phi \wedge \alpha (\grd b {\fff})) \\
    & = \Theta (\alpha \circ f \circ \gamma) .
  \end{align*}
  Since $\alpha\dashv\gamma$, the operation
  $\alpha\circ(-)\circ\gamma:(\Pos(A,A),\le)\to(\Pos(\powo\smems,\powo\smems),\le)$
  is strict and continuous. Therefore
  \begin{align*}
    \asem{\wwhile b P}
    & = \bigvee \Theta^n (\bot)
      = \bigvee \Theta^n (\alpha\circ\bot\circ\gamma)
      \ge \bigvee (\alpha \circ  \Psi^n (\bot)  \circ \gamma)
      = \alpha \circ \left( \bigvee  \Psi^n (\bot)\right)  \circ \gamma\\
    & = \alpha \circ \csem{\wwhile b P} \circ \gamma.
  \end{align*}
\end{proof}
This result is generic with respect to monads, the interpretation of
(effectful) assignment commands, truth values and Galois
connections.
This theorem gives a functorial over-approximation of the best abstract interpretation derived from Galois connections. We see a similar story with the simply typed lambda calculus in \sref{systemt}.
In the following paragraphs, we discuss how \tref{absind} paves the
way to a computable static analysis.

\paragraph{Evaluation of the abstraction of an interpretation
  with a finite height domain.}
When the abstract domain is a complete lattice with the finite chain
property (which asserts that any totally ordered subset of the lattice
is finite), \tref{absind} provides an algorithm to compute the abstract
semantics that it defines~\cite{cc:popl:77}.
Indeed, the complete lattice property and the finite chain property
respectively ensure the definition and the termination of the
computation of \( \asem{\wwhile b P} \).
A classical analysis that falls into this case is the analysis with
the lattice of constants~\cite{gk:popl:73}.
  
\paragraph{Evaluation of the abstraction of an interpretation using
  fixpoint over-approximation.}
In practice, few abstract domains satisfy the properties that were
required in the previous paragraph.
First, the complete lattice property often fails to hold.
Then, the existence of least upper bounds for any family of abstract
elements can then not be guaranteed.
Second, even when any family of abstract elements has a least upper
bound,  the $\omega$-chain generated by
\( \asem{\wwhile b P} \) 
may not converge after finite iterations, hence its least upper
bound may not be computable.
As an example, this occurs in the case of the abstract domain of
intervals~\cite{cc:popl:77}: as it is possible to find an infinite
sequence of intervals \( I_0 \subset I_1 \subset \ldots I_n  \subset
I_{n+1} \subset \ldots \), the definition of \( \asem{\wwhile b P} \)
that was provided earlier does not terminate in general.

Such cases are typically addressed by specific fixpoint approximation
abstract operators.
The most common instance is \emph{widening}~\cite{cc:popl:77} and
consists of a binary operator $\triangledown$ that over-approximates
concrete unions and ensures termination in the sense that any sequence
of applications of widening stabilizes after finitely many iterates.
We now detail how our framework accommodates such analyses.
We use the same notations as in the previous definition of \( \asem{.} \)
and assume a program $P$, a condition $b$, and an abstract element
$\phi$.
Then, we define $W_n^\phi\deq\bigvee^{i=0...n}_A\tau^i(\phi)$ where $\tau(x)\deq\asem P(x\wedge_A \alpha(\grd b \ttt))$.
Interpolation operators such as widening produce an over-approximation
of all the elements of this sequence in terms of $\leq$.
As an example, in the case of widening, the analysis computes a
sequence defined by $V_0^\phi \deq W_0^\phi$ and $V_{n+1}^\phi \deq V_n^\phi\mathbin{\triangledown}
\tau(V_n^\phi)$.
This sequence converges to an element that we note $V^\phi$.
Then the analysis produces $\asem{\wwhile b P}(\phi) \deq V^\phi \wedge_{A}
\alpha(\grd b \fff)$.

\paragraph{Category of abstractions of interpretations.}
We end this section by introducing the category of {\em abstractions}
of an interpretation. This can be viewed as a categorical
understanding of the so-called \emph{lattice of abstractions}.
Fix an interpretation of a category $L$, that is, an oplax functor
$C\colon\,L\to\Pos$. This interpretation would typically be a
\gcs{} as in \sref{3:colsem}. The category of
abstractions of $C$ is the category where an object is a \conci{}
$\{\gamma_a:A(a)\to C(a)\}_{a\in L}$ from an interpretation $A$ to $C$,
and a morphism from $\{\gamma_a:A(a)\to C(a)\}_{a\in L}$ to
$\{\gamma_a':A'(a)\to C(a)\}_{a\in L}$ is a \conci{}
$\{\delta_a:A(a)\to A'(a)\}_{a\in L}$ such that for any object $a$ of $L$,
$\gamma_a'\circ\delta_a \geq \gamma_a$,
saying that $\gamma_a'\circ\delta_a $ is more abstract
than $\gamma_a$. 
We
will denote it by $\Abs{C}$. In words, $\Abs{C}$ is the oplax slice
category over $C$ of the 
category of interpretations of $L$ and concretizations between them.


Some standard constructions in abstract interpretation can be lifted
to categorical structures of $\Abs{C}$, provided that $C$ enjoys
certain properties.  When $C$ factorizes through the category of
meet-semilattices and meet-preserving maps, the category $\Abs{C}$ has
binary products.  It is given by the \emph{Cartesian product abstract
  domain} (see e.g. \cite{Cortesi_2013}).  Concretely, given two
concretizations $\{\gamma_a:A(a)\to C(a)\}_{a\in L}$ and
$\{\gamma_a':A'(a)\to C(a)\}_{a\in L}$, their binary product is given by
the interpretation $A\times A'$ and the concretization
$\{\delta_a:A(a)\times A'(a)\to C(a)\}_{a\in L}$ where
$ \delta_a(u,v) = \gamma_a(u)\wedge\gamma'_a(v)$.  It remains to be
seen what other categorical structures are available on $\Abs{C}$.

\section{Abstracting Denotational Semantics of $\lambda$-Calculus}
\label{s:systemt} \label{s:5:systemt}

In this section, we develop another kind of example to demonstrate the
modularity of our theory of abstractions.
Examples discussed so far only consist of imperative languages with various
semantics and interpretations.
Here, we describe interpretations for the simply typed lambda calculus over
a higher-order signature, as a larger case study to demonstrate that our
theory also copes with functional languages.
In this study, we adopt rather simple {\em non-relational abstraction}
of domains, that is, the abstract domain for product types is the
product of abstract domains for component types.  

\subsection{The Language Category: the Free Cartesian Closed
  Category}

Given a set $B$, we define $Typ(B)$ to be the set of types generated from
$B$ with the type constructors for the unit type $1$, the binary product
type $\times$ and the arrow type $\to$. 
We specify the simply typed lambda calculus by a {\em higher-order signature}. It
consists of a set $B$ of base types, a set $O$ of constants and a
function $typ:O\to Typ(B)$ assigning a type to each constant. We do
not consider equational axioms on constants.
In the rest of \sref{5:systemt}, we let $\Pi$ be a higher-order
signature $(B,O,typ)$.

We write $\lambda(\Pi)$ for the free Cartesian closed category (CCC
for short) generated from $\Pi$ \cite{crole_1994}. An object of
$\lambda(\Pi)$ is a type in $Typ(B)$, and a morphism from $\tau$ to
$\tau'$ is a $\beta\eta$-equivalence class of a term-in-context
$x:\tau\vdash M:\tau'$ in the simply typed lambda calculus over $\Pi$.

\subsection{\GCS{} for the Lambda Calculus}

We next give a \gcs{} for $\lambda(\Pi)$. We first recall the standard
functorial semantics of the simply typed lambda calculus in a
Cartesian closed category (CCC for short).  Let
$(\mcC,1_\mcC,\times_\mcC,\To_\mcC)$ be a CCC.  A {\em
  $\Pi$-structure} in $\mcC$ consists of 1) an assignment
$\sem -_0:B\to\mcC$ of a $\mcC$-object to each base type, and 2) an
assignment $\sem c_0:1_\mcC\to\sem{typ(c)}$ of a $\mcC$-morphism to
each constant $c\in O$; here $\sem-:Typ(B)\to\mcC$ is the inductive
extension of $\sem-_0:B\to\mcC$ by
\begin{align*}
  \sem b&\deq \sem b_0,
  &
    \sem 1&\deq 1_\mcC,
  &
    \sem{\tau_1\times\tau_2}&\deq \sem{\tau_1}\times_\mcC\sem{\tau_2},
  &
    \sem{\tau_1\to\tau_2}&\deq \sem{\tau_1}\To_\mcC\sem{\tau_2}.
\end{align*}
The $\Pi$-structure induces the Cartesian closed functor\footnote{Here
  it means a functor strictly preserving finite products and exponentials.
}
$\sem-:\lambda(\Pi)\to\mcC$ \cite{crole_1994}.

For our \gcs{}, we employ the CCC $\Set$ and fix a $\Pi$-structure in $\Set$,
then take the induced Cartesian closed functor $\sem-$ as a denotational semantics.
For the property functor, we take the covariant powerset functor $Q:\Set\to\Pos$ from
Section~\ref{s:3:colsem}. To summarize, we take the following \gcs{}:
$
Q\circ{\sem-}:
\lambda(\Pi) \to
    \Set \to
    \Pos
$.

\subsection{Abstracting Collecting Semantics by Galois Connections}

Thanks to Theorem~\ref{t:galcon}, to construct an
abstract semantics of $\qpow{\sem{-}}$, it suffices to give a
Galois connection of the form
$\galois{\alpha_\tau}{\gamma_\tau}{A(\tau)}{\qpow{\sem{\tau}}}$ for
every type $\tau\in Typ(B)$. We do so by first assuming, for each {\em
  base} type $b\in B$, 1) a poset $A_0(b)$ of abstract base type
properties, and 2) a Galois connection
$\galois{\alpha_b}{\gamma_b}{A_0(b)}{\qpow{\sem b}}$.  For instance,
when there is a base type $nat\in B$ for natural numbers and it is
interpreted as the set $\Nat$ of natural numbers, we may take the poset of
intervals (including the empty one) over $\Nat$ for $A_0(nat)$, and
take the Galois connection for interval abstraction for
$\galois{\alpha_{nat}}{\gamma_{nat}}{A_0(nat)}{\qpow{\Nat}}$.

We then inductively extend the mapping $A_0:B\to\Pos$ to the one
$A:Typ(B)\to\Pos$ by the Cartesian closed structure of $\Pos$:
\begin{gather*}
  A(b)\deq A_0(b),\quad
  A(1)\deq 1_\Pos,\quad
  A(\tau_1\times\tau_2)\deq A(\tau_1)\timespos A(\tau_2),\quad
  A(\tau_1\to\tau_2)\deq A(\tau_1)\Topos A(\tau_2).
\end{gather*}
We next construct a type-indexed family of Galois connections. Below
$\galois{\alpha_{\tau_i}}{\gamma_{\tau_i}}{A(\tau_i)}{\qpow{\sem{\tau_i}}}$
are Galois connections for $i = 1,2$. (For
$\galois {\alpha_1} {\gamma_1} {1_\Pos} {Q(\sem 1)}$ we take the
unique one.)
\begin{displaymath}
  \xymatrix@C=2cm{
    A(\tau_1\times\tau_2)
    & \qpow{\sem{\tau_1}}\timespos\qpow{\sem{\tau_2}} \adjunction{l}{\alpha_{\tau_1}\times\alpha_{\tau_2}}{\gamma_{\tau_1}\times\gamma_{\tau_2}}
    & \qpow{\sem{\tau_1\times\tau_2}} \adjunction{l}{\alpha^\times}{\gamma^\times} \\
    A(\tau_1\to\tau_2)
    & \qpow{\sem{\tau_1}}\Topos\qpow{\sem{\tau_2}} \adjunction{l}{\alpha_{\tau_2}\circ \_ \circ\gamma_{\tau_1}}{\gamma_{\tau_2}\circ\_\circ\alpha_{\tau_1}}
    & \qpow{\sem{\tau_1\to\tau_2}} \adjunction{l}{\alpha^{\To}}{\gamma^{\To}}
  }
\end{displaymath}
Here, Galois connections $\alpha^\times\dashv \gamma^\times$ and
$\alpha^{\To}\dashv \gamma^{\To}$ are given by
\begin{align*}
  \alpha^\times(U) &\deq (Q(\pi_1)(U),Q(\pi_2)(U)) & \gamma^\times(U,V) &\deq U\times V \\
  \alpha^{\To}(F) &\textstyle\deq \lam{U}\bigcup_{f\in F}Qf(U) & \gamma^{\To}(g) &\deq\{f \mid \forall U \subseteq \sem{\tau_1}.\, Qf(U) \subseteq g(U)\}.
\end{align*}
This establishes a type-indexed family of Galois connections
$G\deq
\{\galois{\alpha_\tau}{\gamma_\tau}{A(\tau)}{Q(\sem\tau)}\}_{\tau\in
  Typ(B)}$.

Using the notations of Theorem~\ref{t:galcon}, we obtain an abstract
interpretation, that is, an oplax functor
$\csem-^G:\lambda(\Pi)\to\Pos$ given by \funcdefi
{\csem{\tau}^G}{A(\tau)} {\csem{x:\tau\vdash
    P:\sigma}^G}{\alpha_\sigma\circ Q(\sem{x:\tau \vdash
    P:\sigma})\circ\gamma_{\tau}.}  By \tref{galcon}, it comes with a
\conci{} $\gamma$ from $\csem-^G$ to $Q\circ\sem-$, as well as an
\absi{} $\alpha$ from $Q\circ\sem-$ to $\csem-^G$.  The interpretation
$\csem-^G$ is normal when the base type Galois connections are Galois
insertions: indeed, it can be easily proved by induction on types that
$\alpha_\tau \circ \gamma_\tau = \id$.  However, it is not functorial.

To have an inductively defined semantics that over-approximates
$\csem-^G$, we derive a new semantics of the lambda calculus that
interprets each constant $c$ by
$\csem{x:1\vdash c:typ(c)}^G:1_\Pos\to A(typ(c))$. Formally, we form a
$\Pi$-structure $(A_0,\{\csem{x:1\vdash c:typ(c)}^G\}_{c\in O})$ in
$\Pos$ and induce a Cartesian closed functor
$\psem-:\lambda(\Pi)\to\Pos$. This process is a lambda-calculus
analogue of the derivation of $\asem-$ from $\csem-^G$ for the while
language in \sref{4:2:gcabsint}; in this example, assignment commands
are replaced with constants in $\Pi$.
The following theorem is in parallel with \tref{absind}.
\begin{theorem}\label{t:absindlam}
  $\psem-$ is a Cartesian closed functor, and $\csem-^G\le\psem-$.
\end{theorem}

\section{Related Works}
In this paper, we study the construction of frameworks to describe semantic
abstraction using categorical tools which are already used heavily to
construct semantics of programming languages.

Abstract interpretation~\cite{cc:popl:77,cc:popl:79} was proposed as a very
general framework to compare program semantics and has found applications in
many areas of computer science such as semantics~\cite{pc:mfps:97}, program
analysis and verification~\cite{cc:popl:77,astree:pldi:03,fb:cacm:19}, program
transformations~\cite{cc:popl:02,xr:popl:04}, or security~\cite{gm:popl:04}.
It is of very general scope as shown in \cite{cc:jlc:92}, since it can be
adapted to various program semantics, styles of abstraction relations
(abstraction functions, concretization functions, Galois connections, or
basic abstraction relations), abstract domains, and classes of approximation
algorithms.
Although it was remarked very early that it could also be formalized
in category theory, most descriptions use set theory.  Steffen et
al~\cite{sjm:tia:92} define both concrete and abstract semantics in a
categorical framework in which they express soundness.  Before their
work, Panangaden and Mishra give a categorical formulation of abstract
interpretation based on the concept called {\em quasi-functor}
\cite{panangaden}. We leave the comparision of quasi-functors with
oplax functors for future work.
Sergey et al.~\cite{might:pldi:13,might:oopsla:15} rely on monads to
construct a control flow analysis of a small functional language, where
various aspects of the program semantics and abstraction are described
by monads.
Our work uses categorical tools in a different way.
First, we use a framework based on oplax functors and lax natural
transformations in order to express and compare
semantics.
This is the right setting for embracing the induction of semantics by Galois connections (\tref{galcon}), which is foundational in abstract interpretation.
Second, we build upon a functorial decomposition to generalize the notion
of collecting semantics, to expose a property component and a
semantic component.
According to~\cite{cc:jlc:92} as \emph{``the main utilization of the
  collecting semantics is to provide a sound and relatively complete
  proof method for the considered class of properties''}, however this
presentation does not give a systematic way to construct one such
semantics, thus one of our contributions is to provide a functorial
interpretation for this notion.
Another of our contributions is to bridge the gap between monadic
tools to construct program semantics and their use in abstract
interpretation.

This paper focuses on the forward property-based semantics.  The other
side of the story, namely the {\em backward} property-based semantics,
has recently been categorically studied
\cite{DBLP:journals/tcs/Hasuo15,DBLP:conf/mfps/AguirreK20,wolter_2022}. They
point out that Dijkstra's {\em weakest precondition predicate
  transformer} (wppt for short), which is a typical backward
property-based semantics, corresponds to functors of type
$L^{op}\to\Pos$ (see $wp^o$ in \eref{pfkleisli}).  This fact suggests
that we may dualize this paper's story to develop abstract
interpretations for backward property-based semantics. There, wppts
expressed as functors would play the role of \gcs{}.


\section{Conclusion}

We have introduced categorical structures that account for basic
concepts and constructions in abstract interpretation.
We have demonstrated that it can account for the abstraction of both
imperative and functional programs in a unified manner.
A future work is to explore various combinations of the data in the
diagram \eqref{eq:abs}.
For the language category $L$, it would be interesting to take
programming languages other than the while language and the simply
typed lambda calculus, such as linear lambda calculus
\cite{10.1007/3-540-56992-8_6}, Moggi's monadic metalanguage
\cite{m:ic:91}, and dependent type theories. For the property functor
$C$, {\em regular fibrations} \cite[Definition 4.2.1]{b:jacobs:00} are
a rich source.
The construction of
$(A,\gamma)$-pairs is the central part of the development of abstract
interpretation, and it remains to be seen if our categorical formalism
provides new construction methods of $(A,\gamma)$-pairs.

\begin{ack}
  The first and third authors were supported by ERATO HASUO
  Metamathematics for Systems Design Project (No. JPMJER1603), JST.
  The third author was also supported by JST, CREST Grant Number JPMJCR22M1.
  The second author was supported by the French ANR VeriAMOS project.
  The authors are grateful to James Haydon for his critical reading of
  the manuscript, and implementation proposals and suggestions of new
  directions during discussions. The authors are also grateful to
  Ichiro Hasuo for insightful comments and suggestions.
\end{ack}

\bibliographystyle{./entics}
\bibliography{mfps}

\end{document}